\documentstyle[aps,preprint]{revtex}

\draft
\begin{document}
\author{Yan V. Fyodorov$^{}$\cite{leave}  and Hans-J\"{u}rgen Sommers}
\address{ Fachbereich Physik,
Universit\"{a}t-Gesamthochschule Essen\\
 Essen 45117, Germany}
\title{"Level Curvature" Distribution for Diffusive Aharonov-Bohm
systems: analytical results}

\date{\today}
\maketitle

\begin{abstract}
We calculate analytically the
distributions of "level curvatures" (LC) (the second derivatives of
eigenvalues with respect to a magnetic flux) for a particle moving

in a white-noise random potential.
 We find that the Zakrzewski-Delande conjecture \cite{Del} is
still valid even if the lowest weak localization corrections are taken
into account. The ratio of mean level curvature modulus to mean dissipative
conductance
is proved to be universal and equal to $2\pi$ in agreement with available
numerical data.
\end{abstract}

\pacs{PACS numbers: 05.45.+b, 71.55.J}

\narrowtext

Nowadays it is considered to be a well established fact that spectral
statistics of generic chaotic or disordered systems is
adequately described by that typical
for eigenvalues of large Gaussian random  matrices (RM)\cite{Bohigas}.
The range of
 applicability of the "Gaussian universality" to real disordered
systems was considered in
the pioneering papers by Efetov\cite{Efrev} and Altshuler and Shklovski
\cite{Altshkl}. It was demonstrated that statistical properties of the energy
levels of a quantum particle moving in a static random potential follow
RM predictions as long as effects of Anderson localization on the particle
diffusion are neglegible. Another important result
 is due to Berry who demonstrated how the same universality
may arise in globally chaotic ballistic systems ( "quantum billiards")
\cite{Berry}.

Quite recently an interesting new developement in the
study of spectra
of disordered systems and their chaotic counterparts has been made.
The problem is to study
 the so-called "level response statistics",
i.e. to provide a statistical description of
sensitivity of the energy levels to external perturbations of different
types. This issue attracted a great deal of research interest, both
analytically and numerically\cite{Wilk}-\cite{Keating},
\cite{Del}-\cite{brmon}.
The most frequently studied characteristics are
first and second derivatives of the energy levels $E_{n}(\alpha)$
with respect to a tunable parameter $\alpha$ characterizing the strength
of perturbation. Physically the role of such a parameter can be played by, e.g.
an external magnetic field, the strength of a scattering potential for
disordered metal, a form of confining potential for quantum billiards,
or any other appropriate parameter on which the system Hamiltonian
is dependent. The first derivatives $v_{n}=\partial E_{n}/\partial \alpha$
are frequently called the "level velocities" (LV)
(or "level currents"),
the second ones $K_{n}=\partial^{2}E_{n}/\partial \alpha^{2}$ are known
as "level curvatures".

In a series of papers by the MIT group\cite{MIT,Jap}, see also \cite{Been},
 it was found that
for a generic chaotic system whose unperturbed spectrum is well described
by the universal RM statistics the set of "level velocities" $v_{n}(\alpha)$
is characterized (after appropriate rescaling) by a universal
correlation function $\langle v_{n}(\alpha)v_{n}(\alpha')\rangle$
whose form is again dependent only on the symmetry
of the unperturbed Hamiltonian
and that of the perturbation.

The range of applicability of these results to real systems is the same as
before: they are valid for systems with completely "ergodic" chaotic
eigenfunctions covering randomly, but uniformly all the available
phase space and
showing no specific internal structure. The universality
is believed to be
independent of "whether the chaos originates in
mesoscopic disorder or deterministic
instability of the classical trajectories."\cite{Keating}.
The effects of eigenfunction localization-- either due to scarring
\cite{Heller} or due to disorder-induced quantum interference
 (the Anderson localization) -- result in substantial
modifications of the LV characteristics, see \cite{Del,my,FM,Takami}

Much interest was concentrated on the "level curvature" (LC) distribution
${\cal P}(K)$. Gaspard and coworkers \cite{Gasp} discovered the
universal asymptotic behaviour ${\cal P}(K)\propto K^{-(\beta+2)}$ in the
large curvature limit $K\to\infty$, the parameter $\beta=1,2$ or $4$
depending on the symmetry universality class. This behaviour is a direct
consequence of the so-called "level repulsion" for unperturbed systems.
Namely, the probability ${\cal P}(s)$ for two unperturbed
energy levels to be separated
by the spacing $s$ much smaller than the mean level spacing $\Delta$
vanishes as ${\cal P}(s)\propto s^{\beta}$, and this fact can be shown to
result in the abovementioned large-curvature behaviour\cite{Braun}.
The analytical form of the whole distribution function ${\cal P}(K)$
was guessed by Zakrzewski and Delande\cite{Del} on the basis of numerical
results and is given by the following expression:\cite{foot2}

\begin{equation}\label{zd}
{\cal P}(k)=C(\beta)(1+k^{2})^{-(1+\beta/2)}
\end{equation}
where $C_{\beta}$ is a normalization constant, $\beta=1,2,4$ depending on
presence or absense of time-reversal invariance and Kramers degeneracy
and $k$ is the dimensionless
level curvature.
We will refer to this expression as to the ZD conjecture.

Very recently, von Oppen\cite{Oppen}
succeeded in demonstrating that eq.(\ref{zd})
is indeed exact for the ensembles of large Gaussian Hermitian
$(\beta=2)$ and real symmetric matrices.
The validity of ZD conjecture for all three
classes of large Gaussian matrices was proved
in a different way by the present authors\cite{FS}.
It is natural to suppose, that the status of
the distribution eq.(\ref{zd})
within the theory of disordered and chaotic systems is the same as before:
it is valid for systems with completely ergodic extended eigenstates.
This was indeed found to be the case in a series of interesting
numerical experiments\cite{Izr,brmon}
 on quasi-1D as well as 3D periodic random
tight-binding models subject to the influence of Aharonov-Bohm magnetic
flux, which
acts as a time-reversal symmetry breaking parameter. In particular, the
distribution eq.(\ref{zd}) was found to persist up to the Anderson localization
transition\cite{Izr}.

Unfortunately, the methods used in \cite{Oppen,FS} for deriving
the form of LC distributions for the Gaussian ensembles
were havily based upon the explicit form of the joint probability
density of eigenvalues known for all these ensembles
\cite{Bohigas}, which is of course
an immense simplification. One therefore has to invent
a different technique in order to be able to treat
the curvature distribution analytically under more realistic assumptions.

It turns out to be possible to find such
a technique for the case of
time-reversal
invariant systems subject to a time-reversal symmetry breaking perturbation.
This case seems to be one of the most interesting from the physical point
of view as well as relevant experimentally. As a physical realisation
one can imagine a disordered mesoscopic sample
(e.g. cylinder or ring) pierced by magnetic
flux $\phi$. For such a system the "typical" level curvature is expected to
be related to the dimensionless conductance $g_{c}$ of the sample
due to the famous Thouless formula:
$ \frac{1}{\Delta}
\partial^{2} E_{n}/\partial \phi^{2}|_{\phi=0}\sim
E_{c}/\Delta\equiv g_{c}$.
Initially suggested by Thouless\cite{Thou}, this relation attracted renewed
interest recently. Its meaning was reconsidered in a broader context
by Akkermans and Montambaux \cite{Ak}, see also a quite detailed discussion in
\cite{Braun,brmon}.
All these facts make the consideration of such systems to be of
special interest.

A specific feature allowing us to treat
the particular case of weak time-reversal symmetry breaking
perturbation analytically
is vanishing of the first derivatives $\partial E_{n}/\partial\phi|_{\phi=0}$
on reasons of symmetry. This fact allows one to represent the
curvature distribution in terms of a product of advanced and retarded Green
functions, the average over the disorder being performed nonperturbatively
with help of Efetov's supersymmetry approach\cite{Efrev}. The method
provides a unique possibility to derive the level
curvature distribution starting from a genuine
microscopic Hamiltonian of a quantum particle experiencing
elastic scattering: \begin{equation}\label{Ham}
{\cal H}(\phi)=\frac{1}{2m}\left(\bbox{p}-\frac{e}{c}\bbox{A}(\phi)
\right)^{2}+U(\bbox{r})
\end{equation}
with $U(\bbox{r})$ being a white noise random potential and $\bbox{A}(\phi)$
standing for the vector potential corresponding to the magnetic flux $\phi$.

 This fact allows one to try to take
into account the weak localization effects due to the finite
ratio of the Thouless energy $E_{c}$ to the mean level spacing $\Delta$.
 The general way of doing this was developed recently by Kravtsov and
Mirlin\cite{KM}.
Exploiting this method one can find that the curvature distribution
eq.(\ref{zd}) preserves its form
 to the first order in $\Delta/E_{c}\ll 1$, the width $\langle |K|\rangle$
being renormalized.

To begin with, we introduce the
resolvent operator
$\hat{G}^{\pm}(\alpha;\epsilon)=
\left[E-{\cal H}(\phi)\pm i\epsilon\right]^{-1}$
where $\alpha=\phi/\phi_0$, with $ \phi_0=2\pi c/e$ being the flux quanta.
Let us now consider the following
correlation function:
\begin{equation}\label{cor}\begin{array}{c}
{\cal K}(u)=\\
\lim_{\epsilon\to 0}\epsilon
\mbox{Tr}\hat{G}^{+}(\alpha=2\sqrt{\epsilon/u};\epsilon)
\mbox{Tr} \hat{G}^{-}(\alpha=0;\epsilon)\equiv \\
\lim_{\epsilon\to 0}\sum_{n,m=1}^{N} \frac{\epsilon}
{[E-E_{n}(\alpha=0)-i
\epsilon][E-E_{m}(\alpha=2\sqrt{\epsilon/u})+i\epsilon]}
\end{array}\end{equation}
Here we used the expression for the trace
of the resolvent $\mbox{Tr}\hat{G}$ in terms
of eigenvalues of the Hamiltonian $\hat{H}(\phi)$.

It is important to note that the limiting procedure $\epsilon\to 0$ in eq.
(\ref{cor}) is performed prior to the
thermodynamic limit $V\to \infty$, where
$V$ is the system volume. Therefore,
the parameter $\epsilon$ (that plays a role of effective "level broadening"
necessary to regularize the resolvent operator) can be considered as small in
comparison with the mean level spacing $\Delta\propto 1/V$. It was already
mentioned that the
probability of having two levels at a distance $s\ll \Delta$ is vanishingly
small due to level repulsion. Taking this fact into account one finds that
 the only terms that survive the limiting
procedure $\epsilon \to 0$ are those with coinciding indices $m=n$.
Remembering $\partial E_{n}/\partial\phi|_{\phi=0}$, one obtains:
\begin{equation}\label{curv}
{\cal K}(u)=\pi\sum_{n}\frac{u^2+iu\,K_{n}}{u^2+K_{n}^{2}}\delta(E-E_{n});
\quad K_{n}=\frac{\partial^2 E_{n}}{\partial {\alpha}^2}|_{\alpha=0}
\end{equation}
Let us perform formally
the averaging over the disorder and introduce the function
${\cal
P}(K)=\left\langle\Delta
\sum_{n=1}^{N}\delta(K-K_{n})\delta(E-E_{n})\right\rangle$. This function
has the meaning of distribution of curvatures
for levels in a narrow spectral window around the  energy $E$ .
Then one obtains the following relation:
\begin{equation}\label{main}
\frac{\Delta}{\pi}\langle {\cal K}(u)\rangle= u\int_{0}^{\infty}
d\,v e^{-u\,v}\int_{-\infty}^{\infty} dK\,{\cal P}(K)\exp{(i\,v K)}
\end{equation}

Let us stress that it is simultaneous
vanishing of all
the first derivatives  $v_{n}=\partial E_{n}/\partial \phi$
at $\phi=0$  that made it possible to scale
$\phi/\phi_0\propto \epsilon^{1/2}$ producing
the finite limiting expression eq.(\ref{curv})
when $\epsilon\to 0$ in eq.(\ref{cor}).
In contrast, when $v_{n}\ne 0$ one can
scale $\phi\propto \epsilon$ and get
 the expression eq.(\ref{main}), but with
the {\it level velocity} distribution ${\cal P}(v)$ substituted for the
{\it level curvature} distribution ${\cal P}(K)$. This fact was already
used in \cite{my,Jap} in order to calculate
analytically LV distributions for different systems.

In order to evaluate the average correlation
function ${\cal K}(u)$ one can use Efetov's
 supersymmetry approach\cite{Efrev}.
Exploiting the weak disorder parameter
$k_Fl\gg 1$, with $k_F$ denoting the Fermi
wavenumber and	$l$ denoting the mean free path due to elastic
scattering, the problem can be mapped onto the so-called nonlinear graded
$\sigma$-- model. Rather detailed exposition of the mapping can be found, e.g.
in \cite{Alt}. The final expressions appeared frequently in literature,
most recently in \cite{Fal}, and can be used for our needs after a slight
modification resulting from the fact that in our case the magnetic flux
enters only one of two Green functions rather than both as in \cite{Alt,Fal}.
As the result, one finds:
\begin{equation}\label{coset}
\begin{array}{c}
\left\langle {\cal K}(u)\right
\rangle=
\frac{\pi^2}{4}\lim_{\epsilon\to 0} \frac{\epsilon}{\Delta^2}
\int {\cal D}\mu(Q)\times \\
 \int \frac{d\bbox{r}}{V}
\mbox{Str}\left[\hat{Q}(\bbox{r})\hat{k}\frac{1+\Lambda}{2}\right]
\int \frac{d\bbox{r}}{V}\mbox{Str}\left[\hat{Q}
(\bbox{r})\hat{k}\frac{1-\Lambda}{2}\right]
\exp{-{\cal S}_{\epsilon}(Q)}\\ {\cal S}_{\epsilon}(Q)=\\
-\frac{\pi D}{8\Delta
}\int\frac{d\bbox{r}}{V}\mbox{Str}\left( \nabla Q-
\frac{e}{c}\bbox{A}_{\epsilon}\left[\hat{Q},\hat{\tau}
\right]\right)^2-\frac{\pi\epsilon}{2\Delta}\int \frac{d\bbox{r}}{V}
\mbox{Str}\hat{Q}\hat{\Lambda}\end{array}
\end{equation}
where $D$ is the classical
diffusion constant $D\propto v_Fl$
due to random scattering , $\bbox{A}_{\epsilon}$
is the vector potential corresponding to the magnetic
 flux through the sample $\phi/\phi_0=2\sqrt{\epsilon/u}$
in the gauge ensuring that $\bbox{A}$ is tangential to the surface of sample;
e.g. $\frac{e}{c}\bbox{A}=\frac{2\pi}{L}\frac{\phi}{\phi_0}\bbox{e}_{\tau}$ for
the ring
geometry.
The notation $[\,\,,\,\,]$ stands for the matrix commutator and $\mbox{Str}$
stands for the
graded trace.
The position dependent $8\times 8$ supermatrices $\hat{Q}(\bbox{r})$ satisfy
the constraint $\mbox{Str}\hat{Q}^2$=1 and can be parametrized as follows:
$\hat{Q}=\hat{T}^{-1}\Lambda
\hat{T}$, where $\hat{T}$ belongs to a graded coset space $\mbox{UOSP}(2,2/4)
/\mbox{UOSP}(2/2)\times \mbox{UOSP}(2/,2)$.
 Other $8\times 8$ supermatrices entering
the eq.(\ref{coset}) are diagonal: $$\begin{array}{cc}
\hat{\Lambda}=\mbox{diag}(I_2,I_2,-I_{2},-I_{2}) &
\hat{\tau}=\mbox{diag}(\hat{\sigma},\hat{\sigma},0,0,0,0)\\
\hat{k}=\mbox{diag}(I_{2},-I_{2},I_{2},-I_{2}) & \hat{\sigma}=
\mbox{diag}(1,-1);\quad I_2=\mbox{diag}(1,1)\end{array}$$

Let us now recall that there are two relevant energy scales for a quantum
particle diffusing
in a closed disordered sample of the size $L>l$

: the mean level spacing $\Delta$ and the Thouless energy $E_c=D/L^2$.
For $\Delta \ll E_c$ it is known \cite{Efrev} that

the main contribution to the functional integral,
 eq.(\ref{coset}) comes from spatially uniform configurations
( a so-called "zero mode"): $\hat{Q}(\bbox{r})=\hat{Q}_0\equiv \hat{T}^{-1}
\hat{\Lambda}\hat{T}$. The condition $g_c=E_c/\Delta\gg 1$ determines the
regime where
statistical characteristics both of eigenfunctions and energy levels of the
Hamiltonian
eq.(\ref{Ham}) are adequately described by classical

RM theory \cite{Efrev,Altshkl,FM}. The corrections to RM results have a form of
regular expansion in the small parameter $g_c^{-1}$. A systematic way to
construct
such an expansion was originally suggested in \cite{KM} and discussed in more
details in
\cite{FM}. The general procedure is as follows: one decomposes the matrix
$\hat{Q}(\bbox{r})$ as $\hat{Q}(\bbox{r})=\hat{T}_0^{-1}\tilde{Q}(\bbox{r})
\hat{T}_0$ and uses the fact that supermatrices $\tilde{Q}(\bbox{r})$ only
weakly fluctuate around the value $\tilde{Q}=\hat{\Lambda}$ as long as $g_c\gg
1$. As
the result, the integration over $\tilde{Q}(\bbox{r})$
in eq.(\ref{coset}) can be performed perturbatively. After that one obtains
a renormalized expression for the "zero mode effective action"
$S_{\epsilon}(\hat{Q}_0)$
governing the integral over spatially uniform configuration $\hat{Q}_0$.
This last integration has to be performed nonperturbatively.

Performing the "perturbative" part of the evaluation of the functional

integral to the first order in the small parameter $g_c^{-1}$
and keeping only terms relevant at $\epsilon\to 0$
one obtains for the quantity $\langle {\cal K}(u)\rangle$ the same expression
 eq,(\ref{coset}) with the following replacement $\hat{Q}(\bbox{r})\to
\hat{Q}_0,\quad S_{\epsilon}(\hat{Q})\to S_{\epsilon}(\hat{Q}_0)$, where
\begin{equation}\label{action}\begin{array}{c}
S_{\epsilon}(\hat{Q}_0)=\\

-\frac{\pi\epsilon}{2\Delta}\mbox{Str}
(\hat{Q}_{0}\hat{\Lambda})-\frac{\pi\epsilon}{u\Delta}b
E_c\left( 1-\int\frac{dr}{V}
\Pi(\bbox{r,r})\right)\mbox{Str}(\hat{Q}_0\hat{\tau}\hat{Q}_0\hat{\tau})
\end{array}\end{equation}
Here we introduced  a sample-dependent geometrical factor $b=
\left(\frac{u}{4\epsilon}\right)L^2\int\frac{dr}{V}
\left(\frac{e}{c} \bbox{A}_{\epsilon}\right)^2$, see

\cite{Fal}, equal for the ring geometry to $b=4\pi^2$.
 In the expression eq.(\ref{action}) the quantity

$\Pi (\bbox{r,r'})$ stands for  the so called diffusion propagator
whose Fourier transform is equal to
$\frac{1}{\pi\nu }\frac{1}{D\bbox{q}^2+\epsilon}$, momentum $\bbox{q}$ going
over all {\it nonzero} values allowed from the quantization condition
in a given sample geometry, $\nu$ being the mean density of states at Fermi
level.

In order to evaluate the integral over $\hat{Q}_0$ nonperturbatively
one has to use the so-called Efetov parametrisation of the corresponding
coset space. A useful discussion of necessary technical details can be found,
in
particular, in the paper \cite{Alt4}.
 Below we present only major steps of the evaluation procedure relegating
most of details to more extended publication \cite{FS}.

After standard manipulations, the integration over the graded coset space
is reduced to performing the following three-fold integral:
\begin{equation}\begin{array}{c}
\frac{\Delta}{\pi^2 }\left\langle K(u)\right\rangle=\\
\lim_{\tilde{\epsilon}
\to 0}\tilde{\epsilon}\int_{-1}^{1} d\lambda
\int_{1}^{\infty}\int_{1}^{\infty} d\lambda_{1}d\lambda_{2}
\frac{(1-\lambda^2)\left(\lambda-\lambda_{1}\lambda_{2}\right)^2}
{\left[\lambda_1^2+\lambda_2^2+\lambda^2-2\lambda\lambda_{1}\lambda_{2}-1
\right]^2}\times \\
\left[1-\frac{2\tilde{\epsilon}}{\tilde{u}}
\left(\lambda_1^2+\lambda_2^2+\lambda^2-2\lambda\lambda_{1}\lambda_{2}-1
\right)\right]\times \\
\exp{\left[-2\pi\tilde{\epsilon}
(\lambda_1\lambda_2-\lambda)-
\pi^2\frac{\tilde{\epsilon}}{\tilde{u}}
\left(\lambda_1^2+\lambda_2^2-\lambda^2-1\right)\right]}
\end{array}\end{equation}
where we introduced the notations $\tilde{u}=u
\left(\frac{2bE_c}{\pi}[1-\int\frac{dr}{V}\Pi(\bbox{r,r})]\right)^{-1},\quad
\tilde{\epsilon}=\frac{\epsilon}{\Delta}$.

A close inspection of this expression makes it clear that the only contribution
nonvanishing in the limit $\tilde{\epsilon}\to 0$ is that coming from the
region $\lambda_1\sim\lambda_2\propto \tilde{\epsilon}^{-1/2}$.
Picking up such a contribution one gets after some algebra:

\begin{equation}\label{interm}\begin{array}{c}
\frac{\Delta}{\pi}\left\langle K(u)\right\rangle=\\
\frac{\tilde{u}}{2\pi}
\int_{0}^{\infty} dv v e^{-\frac{2 \tilde{u}}{\pi}v}
\int_{0}^{\infty}\frac{dy}{y}e^{-v\left(y+y^{-1}\right)}
\int_{-1}^{1}d\lambda\frac{\left[2(1-\lambda^2)+
\frac{\lambda}{v}\right]}{\left(y+y^{-1}-2\lambda\right)}
\end{array}\end{equation}
Two last integrations in this expression can be performed analytically.
Comparing the result with the relation eq.(\ref{main}) one obtains the
following expression
for the Fourier transform of the LC distribution:
\begin{eqnarray}\label{width}
\int_{-\infty}^{\infty}{\cal P}(K)\exp{ivK} dK=\frac{\gamma}{2}
vK_{1}(\gamma v), \\ \gamma=2bE_c\left(1-\int\frac{dr}{V}
\Pi(\bbox{r,r})\right) \end{eqnarray}
where $K_1(z)$ is the Macdonald function.
Inverting this Fourier transform one finally arrives at the
desired "level curvature" distribution:
\begin{equation}\label{fin}
{\cal P}(K)=\frac{1}{2}\frac{\gamma^2}
{\left[K^2+
\gamma^2\right]^{3/2}}
\end{equation}

Thus, we proved that the level curvature distribution for weakly disordered
Aharonov-Bohm systems follows

{\it exactly} the form suggested by Zakrzewski and Delande
 \cite{Del} even beyond the domain of validity of random matrix theory, i.e.
when the lowest weak localization corrections are taken into account.
This conclusion is in good coordination with available data
from recent papers \cite{Izr,brmon} where numerical simulations of
lattice analogues of disordered Aharonov-Bohm systems were performed.
Having in mind a connection betwen asymptotic behaviour of LC distribution
and degree of level repulsion, it is worth mentioning that our result
is also in agreement with the recent paper \cite{KM} where the two-level
correlator was shown to be affected by weak localization corrections
only at the second order with respect to $g_c^{-1}$.

Another interesting point is that our exact results when combined with the
earlier
findings\cite{MIT,Jap}
allow one to prove the universality of the ratio of the LC mean scaled modulus
$\langle
\mid K\mid \rangle/\Delta=\gamma/\Delta$
to the sample dissipative conductance\cite{MIT,brmon,Ak} defined as
$C(0)=\frac{1}{\Delta^2}\overline{
\left\langle \left(\frac{\partial E_n}{\partial
\alpha}\right)^2\right\rangle}$, where the
average goes both over the
flux value and over the disorder. The calculation done in \cite{MIT}
produces the value $C(0)= b E_c/(\pi\Delta)$ for any Aharonov-Bohm
diffusive system when $g_c\gg 1$. Comparing this result with eq.(11) one finds
 $\frac{\gamma}{\Delta C(0)}\mid_{g_c\gg 1}=2\pi$
in a good agreement with the
value $6.7\pm 10\%$ found
in numerical simulations \cite{brmon}.

Finally, let us mention that our method allows one to extend the validity
of the ZD conjecture to large symmetric RM with independent {\it arbitrarily}

distributed entries having finite variance. In particular, the most interesting
ensemble of large {\it sparse} RM \cite{sparse} can be treated analytically in
this way.
These results will be published elsewhere\cite{FS}.

Authors are very obliged to F.Haake for stimulating their interest in
the work and to A.Mirlin and K.\.Zyczkowski

for useful discussions. Y.V.F
is grateful to E.Akkermans, Y.Gefen, F.Izrailev and A.Kamenev for discussions
and to F.
v.Oppen and  G.Montambaux for
communicating him their results prior to publication.

The financial support by SFB 237 "Unordnung und grosse Fluctuationen"
is acknowledged with thanks.


\begin{thebibliography}{99}
\bibitem{foot2}In the paper \cite{Braun} the function ${\cal P}(K)$ was found
analyticlly for $2\times 2$ random matrices. The result is different from
eq.(1), but the asymptotic large curvature behaviour is the same.
\end{thebibliography}
\end{document}